\documentstyle[11pt,newpasp,twoside,epsf]{article}
\def\edcomment#1{\iffalse\marginpar{\raggedright\sl#1\/}\else\relax\fi}
\reversemarginpar

\begin{document}
\title{What can exoplanets tell us about our Solar System?}

\author{Charles H. Lineweaver, Daniel Grether \& Marton Hidas}
\affil{School of Physics, University of New South Wales and the\\
 Australian Centre for Astrobiology, Sydney, Australia\\
charley@bat.phys.unsw.edu.au}

\begin{abstract}
We update our analysis of recent exoplanet data that gives us a partial 
answer to the question: How does our Solar System compare to the other planetary systems in the Universe?
Exoplanets detected between January and August 2002 strengthen
the conclusion that Jupiter is a typical massive planet rather than an outlier.
The trends in detected exoplanets do not rule out the 
hypothesis that our Solar System is typical. They support it.
\end{abstract}
\section{Identifiable Trends in Exoplanet Data}
Despite the fact that massive planets are easier to detect, the mass distribution 
of detected planets is strongly peaked toward the lowest detectable masses.  And despite the 
fact that short period planets are easier to detect, the period distribution is strongly 
peaked toward the longest detectable periods. 
In Lineweaver \& Grether (2002, hereafter LG) we quantified these trends as accurately as 
possible. Here we update this analysis by including the 27 exoplanets detected between
January and August 2002.
As in LG, we identify a less-biased subsample of exoplanets (thick rectangle of Fig. 1).  
Within this subsample, we correct for completeness and then quantify trends in mass and period (Fig. 2)
that are less biased than trends based on the full sample of exoplanets. Straightforward 
extrapolations of these trends, into the area of parameter space occupied by Jupiter, indicates that 
Jupiter lies in a region densely occupied by exoplanets.  

Naef et al. (2001) point out that none of the planetary companions detected so far resembles the giants of 
the Solar System. However, this observational fact is consistent with the idea that our Solar System 
is a typical planetary system.  
Fig. 1 shows that selection effects can easily explain the lack of detections of Jupiter-like planets. 
Exoplanets detected to date can not resemble the planets of our Solar System because the Doppler technique used to 
detect exoplanets has not been sensitive enough to detect Jupiter-like planets.
We may be sampling the tail of a distribution -- the only part that we are capable of sampling. If the Sun were a 
target star in one of the Doppler surveys, no planet would have been detected around it. 
This situation is about to change.

Our analysis suggests that Jupiter is more typical than indicated by previous analyses, including our own (LG). 
For example, in Fig. 2, our $\alpha = -1.6$ slope is slightly steeper than the $\alpha = -1.5$ 
found in LG and is steeper than the $\alpha \sim -1.0$
%
%%%%%%%%%%%%%%%%%%%%%%%%%%%%%%%%%%%%%%%%%%%%%%%%%%%%%%%%%%%%%%%%%%%%%%%%%%%%%%%%%%%%%%%%%%%%%%%%%%%%%%%%%%%%%%%%%%
\clearpage
\begin{figure}[h,t!]
%\epsscale{1.10}
%\plotone{EP_m_p1.eps}
\plotone{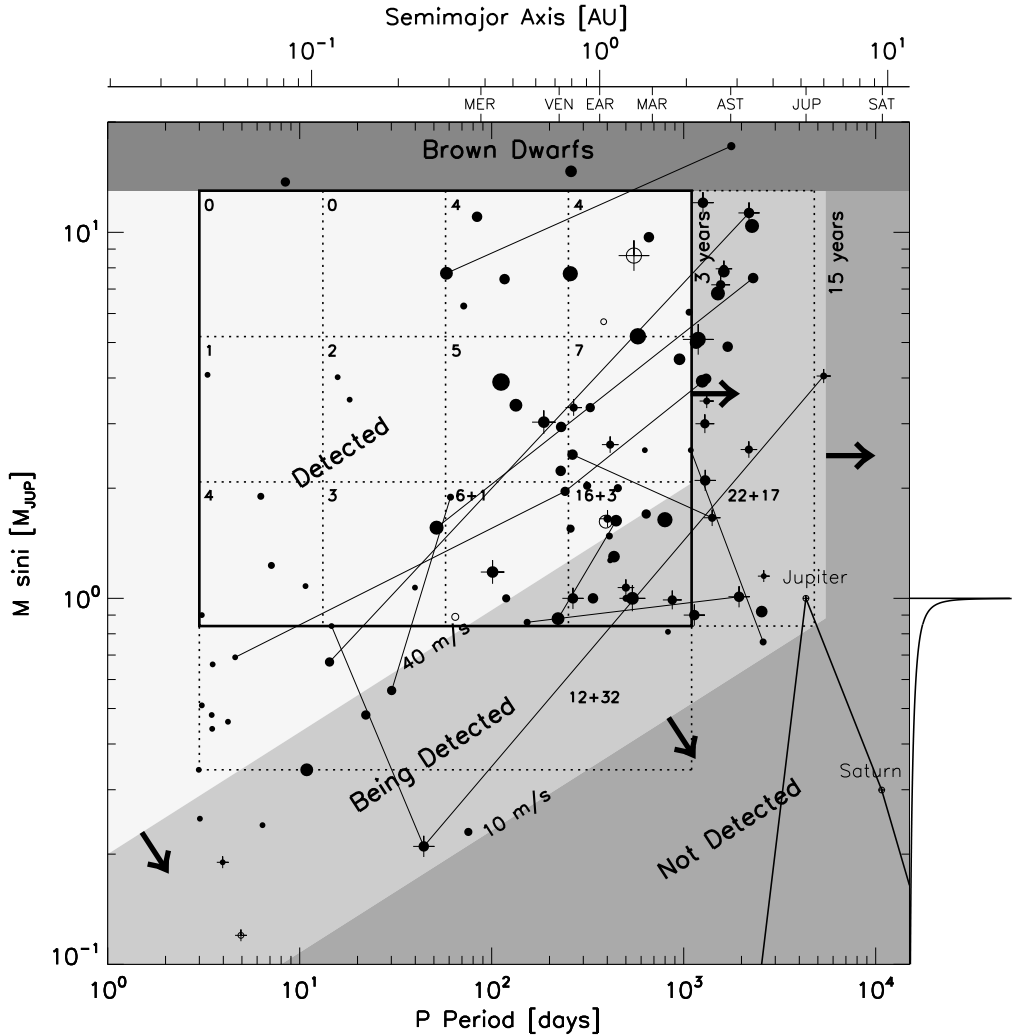}    %EP_m_p2.eps}
%\plottwo{massperiodesp12.eps}{Fig6a_planet_m_p.eps}
%\plotone{massperiodesp12.eps}
%\plotfiddle{massperiodesp12.eps}{8.8 cm}{0}{60}{50}{-180 pts}{-60 pts}
%\plotfiddle {file}{vsize}{rot}{horizontal scale %}{vertical scale %}{horizon translation}{vertical translation}
\caption{
Mass as a function of period for the 101 exoplanets detected as of August 2002. 
Regions where planets are ``Detected'', ``Being Detected'' and ``Not Detected'' by the Doppler surveys 
are shaded differently and represent the observational selection effects of the Doppler reflex technique.  
To identify trends in mass or period we first identify a subsample of these planets that is less biased 
by selection effects. The thick rectangle enclosing the grid of twelve boxes defines such a subsample of 52 planets. 
The number in the upper left of each box gives the number of planets in that box. The increasing numbers 
from left to right and from top to bottom are easily identified trends. 
The two boxes in the lower right lie partially in the ``Being Detected'' region. Thus, they are partially 
undersampled compared to the other boxes within the rectangle.  We correct for this undersampling by 
making the simple assumption that the detection efficiency is linear in the ``Being Detected'' region.
The ``+1'' and ``+3'' in these boxes are the resultant undersampling corrections.  
The ten exoplanetary systems are connected by thin lines.
Open circles indicate detected exoplanets that do not qualify for our sample since they have 
not been monitored for more than 3 years.  
Jupiter is on the edge of the ``Being Detected'' region while Saturn is in the ``Not Detected'' region (right).
Exoplanets detected between January and August 2002 are marked with crosses.
For more details see LG (2002).} 
\end{figure}
\clearpage
\noindent of other previous analyses 
(Jorissen, Mayor \& Udry 2001,  Zucker \& Mazeh 2002, Tabachnik \& Tremaine 2001). This means  that 
(within the same period range) instead of $M_{Jup}$ mass exoplanets being twice as common as  $2 M_{Jup}$ 
exoplanets, we find they are slightly more than three times as common. Similarly we find there are 
$\sim 3$ times as many $0.5 M_{Jup}$ as $M_{Jup}$ exoplanets. When the histogram of all 98 (= 101- 3 brown dwarfs) 
exoplanets is fit, including the highly under-sampled lowest $M sin(i)$ bin, the result is 
$\alpha = -1.2 \pm 0.2$. 
This should be compared to the $\sim -0.8$ of earlier fits (LG) and can be found in Marcy et al. 2002.
This value of $\alpha$  is
close to the $\approx -0.8 \pm 0.2$ found for very low mass stars (Bejar et al. 2001).
When the lowest exoplanet $Msin(i)$ bin is ignored because of known incompleteness we obtain 
$\alpha = -1.5$ (LG reported $-1.1$ for this case). 
Fitting the $M sin(i)$ histogram of the less-biased sample of  52 exoplanets, uncorrected for under-sampling, 
yields $\alpha = -1.5$ (LG reported $-1.3$).  After correcting for under-sampling (with a 4 planet correction) we obtain 
our final result: $\alpha = -1.6 \pm 0.2$ (LG reported $-1.5$).
The 27 exoplanets detected between January and August 2002 fit the trends quantified in LG  but also 
indicate that the slopes are even slightly steeper and thus 
that Jupiters are slightly  more common than indicated by LG.

The biggest uncertainty in this analysis is not the linear approximation used to
make the necessary completeness correction
-- it is probably not knowing how far one can reasonably extrapolate the trends identified.
The surface density of the material in the protoplanetary disk available to 
make planets has to decrease and then drop off at some point. Thus,
there is a danger of extrapolating a trend into this region -- but where is it?
This uncertainty is why we did not extend our analysis to Saturn's orbit and declare 
more speculatively that Saturns are typical gaseous planets.
However, since the exoplanet data indicates that Jupiters are common and we know of no models
in which Jupiters are formed readily but Saturns are not, we see no reason to believe 
that this eventual drop off is near the region (inner edge of the ice zone between 4 and 10 AU) 
where abundant material is expected.
Thus, since the data indicates that Jupiters are abundant, the most reasonable hypothesis is that Saturns 
probably are too. Exoplanet data on Saturns ( $\sim 29$ year orbital period) may be available 
in a decade or two.
%

%%%%%%%%%%%%%%%%%%%%%%%%%%%%%%%%%%%%%%%%%%%%%%%%%%%%%%%%%%%%%%%%%%%%%%%%%%%%%%%%%%%%%%%%%%%%%%%%%%%%%%%%%%%%%%%%%%
%\clearpage
\begin{figure}[h,b!]
%\epsscale{1.1}
%\plottwo{EP_Hist_m2.eps}{EP_Hist_p_extra.eps}
\plottwo{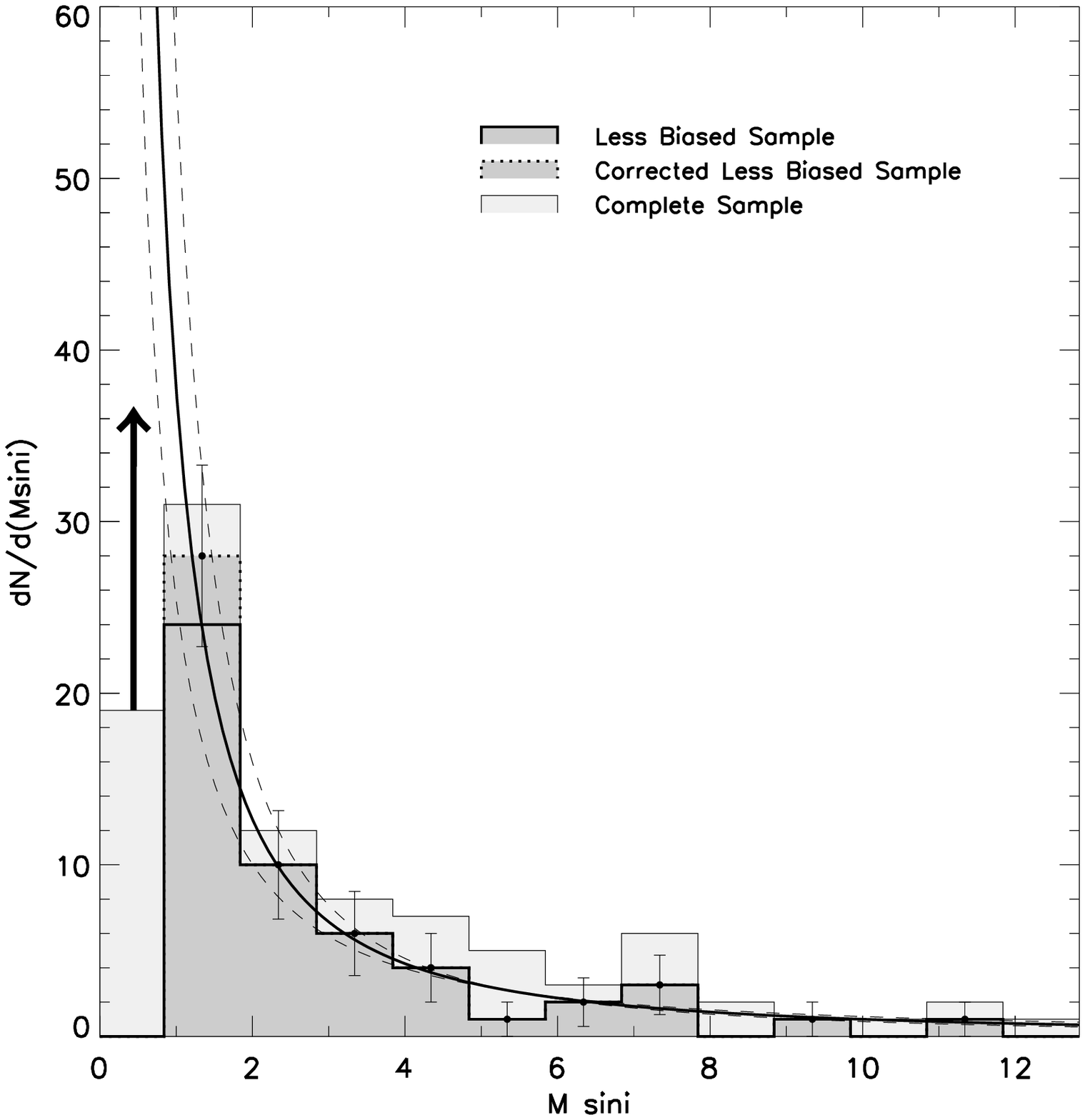}{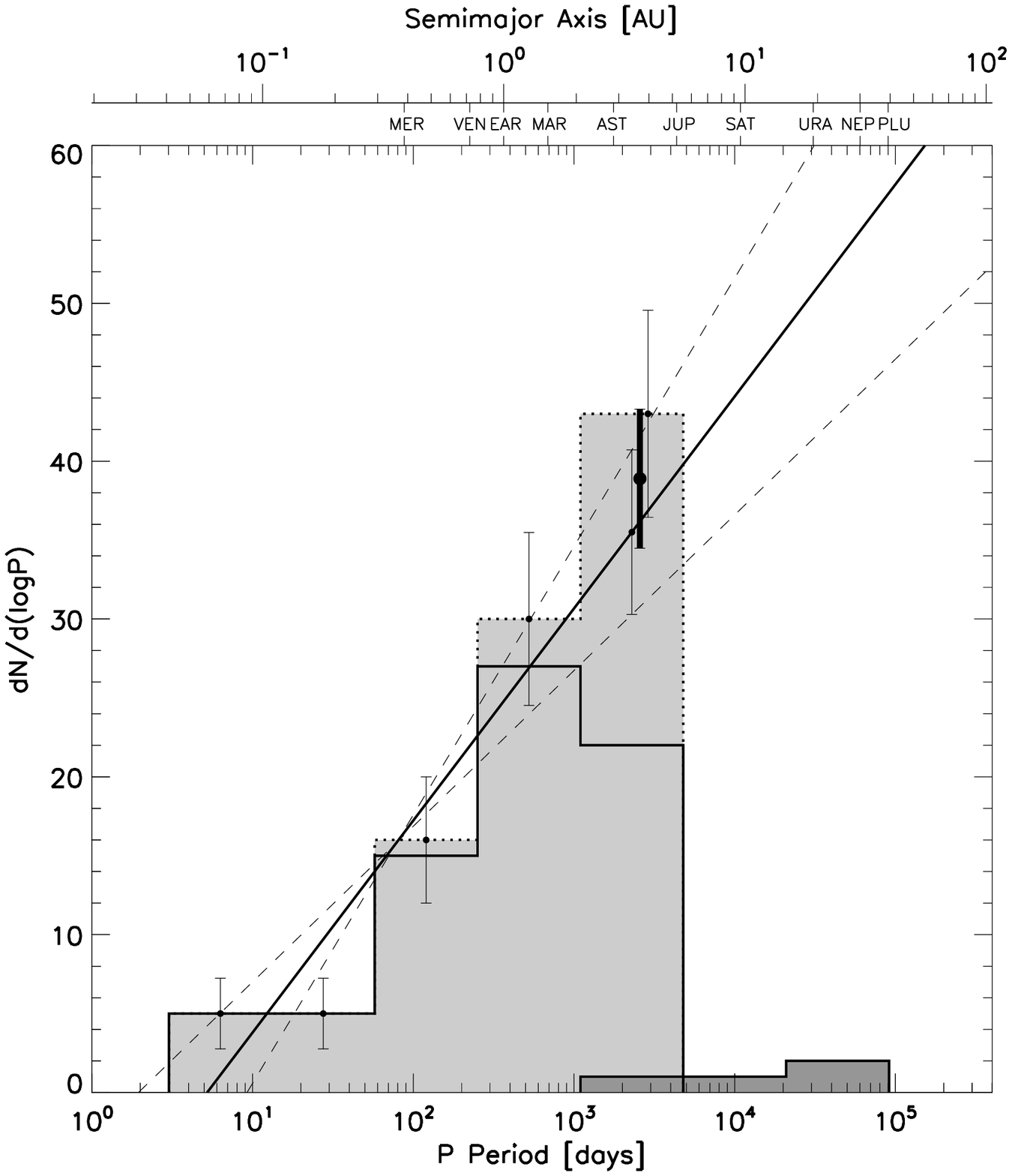}
%\plottwo{masshistogramesp12.eps}{periodhistogramfit10.eps}
%\plottwo{masshistogramesp12.eps}{Fig4extra1.eps}
%\plotone{masshistogramesp12.eps}
\caption{Histogram of exoplanet masses (left). The less-biased subsample of 52 exoplanets 
within the thick rectangle in Fig. 1 is compared here to the histogram of the complete sample 
of 98 (= 101 - 3 brown dwarfs) exoplanets. 
%The errors on the bin heights are Poissonian. 
The solid curve and the enclosing dashed curves are the best fit and 68\% confidence 
levels from fitting the functional form $dN/d Msin(i) \propto Msini^{\alpha}$  to the 
histogram of the corrected less-biased subsample ($56 = 52 + 4$ exoplanets). The extrapolation 
of this curve into the lower mass bin produces an estimate of the substantial incompleteness 
of this bin (arrow). We find $\alpha = -1.6 \pm 0.2$ compared to the $\alpha = -1.5 \pm 0.2$
reported in LG.  
The histograms on the right show the trends in period of the corrected (dotted) and uncorrected (solid) less-biased 
subsample. The line is the best fit to the corrected histogram. The functional form fitted is linear in log P: 
$dN/d(log P) = a log P + b$. 
The best fit slope is $a=13 \pm 4$ slightly steeper than the $a=12 \pm 3$ found in LG. 
We estimate the number in the longest period bin ($1000 < P < 5000$ days) in 
two independent ways: 1) based on the extrapolation of the linear fit and 2) correcting for undersampling in 
the ``Being Detected'' region (Fig. 1). The former yields $36 \pm 5$ while the later yields $43 \pm 7$. 
We take the weighted 
mean of these, $39 \pm 4$, as our best prediction for how many planets will be found in this longest period bin 
scattered over the mass range $0.84 < Msin(i)/ M_{Jup} < 13$. To date, 22 extrasolar planets have been found in 
this period bin. Thus we predict that $17 \pm 4$ more planets will be found in this period bin.
Jupiter, Saturn, Uranus and Neptune are represented by the dark histogram.}
\end{figure}

\end{document}